\documentclass[aps,pra,twocolumn,groupedaddress]{revtex4-2}

\usepackage{graphics}
\usepackage{graphicx}

\begin{document}

\title{Generic quartic solitons in optical media}

\author{Eduard N. Tsoy}
\email[]{e.n.tsoy@gmail.com}
\affiliation{Physical-Technical Institute of the Uzbek Academy of Sciences,\\
2-B, Ch. Aytmatov Str., Tashkent 100084, Uzbekistan}

\author{Laziz A. Suyunov}
\affiliation{Karshi State University, 17, Kuchabag Str., Karshi 180119, Uzbekistan}

\date{26 January 2024}

\begin{abstract}
  Our analysis suggests strongly that stationary pulses exist in nonlinear media with second-,
third-, and fourth-order dispersion. A theory, based on the variational approach, is developed for
finding approximate parameters of such solitons. It is obtained that the soliton velocity in the
retarded reference frame can be different from the inverse of the group velocity of linear waves.
It is shown that the interaction of the pulse spectrum with that of linear waves can affect the
existence of stationary solitons. These theoretical results are supported by numerical simulations.
Transformations between solitons of different systems are derived. A generalization for solitons
in media with the highest even-order dispersion is suggested.
\end{abstract}


\maketitle


\section{Introduction}
\label{sec:intro}

   A balance between second-order dispersion, or group velocity dispersion (GVD), and
cubic (Kerr) nonlinearity results in a formation of optical solitons -- stable pulses that
propagate without change of parameters~\cite{Agra07}. Moreover, these pulses preserve their
shapes and parameters after interaction with each other. Additional effects, such as
higher-order dispersion, the Raman frequency shift, and self-steepening, change parameters
of solitons, see e.g. Ref.~\cite{Agra07}.

   Recently, it was found that stable pulses exist also in media with quartic dispersion only and Kerr
nonlinearity~\cite{Blan16}. These pulses are called ``pure-quartic solitons'' (PQS). Such
solitons have been studied theoretically and experimentally in several papers, see e.g.
Refs.~\cite{Blan16,Krug18,Tahe19,Tam19,Rung20,Melc20,Widj21,Li22}. Stationary and dynamical properties of PQS
were presented in Ref~\cite{Tam19}. In particular, it was shown that PQS have oscillating
tails, see also Ref.~\cite{Akhm94}. A realization of a laser on PQS was suggested in
Ref.~\cite{Rung20}. The dynamics of cavity solitons in pure-quartic media was considered in
Refs.~\cite{Tahe19,Melc20}.

   In the present paper, we consider general quartic media described by the second-, third- (TOD),
and fourth-order dispersion (FOD) terms. We demonstrate that such media also admits the
propagation of stable localized pulses. Since all dispersion terms are involved, we call
these pulses ``generic quartic solitons'' (GQS) to distinguish them from PQS.
Parameters of stationary solitons are found approximately, using the variational approach.
Regions of the GQS existence in the space of the system parameters are obtained. A relation
between moving solitons in general quartic media and pure quartic media is established. A
generalization of results to higher-order dispersion is discussed.

\section{Model and stationary solitons}
\label{sec:mod}

   The dynamics of optical pulses in nonlinear dispersive media is described by the
modified nonlinear Schr\"odinger (NLS) equation~\cite{Agra07}:
\begin{equation}
  i \psi_z  - \frac{\beta_2}{2} \psi_{\tau \tau} - i \frac{\beta_3}{6} \psi_{\tau \tau \tau} +
  \frac{\beta_4}{24} \psi_{\tau \tau \tau \tau}  + \gamma |\psi|^2 \psi = 0,
\label{gnlse}
\end{equation}
where $\psi(\tau, z)$ is the envelope of the electric field, $\tau$ is the time in the
retarded frame, $z$ is the propagation distance, $\beta_{j}$ is the parameter of dispersion
of the $j$-th order, $j = 2, 3$, and $4$, $\gamma$ is the Kerr nonlinearity parameter. We
consider dispersion terms of up to the fourth order only. We mention that at
$\beta_3 = \beta_4 = 0$, the standard NLS equation is completely integrable~\cite{Zakh72}, and has
the soliton solution. At $\beta_3 = 0$, there is also an exact soliton solution~\cite{Karl94}.
Soliton solutions of Eq.~(\ref{gnlse}) for some sets of parameters $\beta_j$ are found in Ref.~\cite{Krug18}.
These solutions have smooth, non-oscillating tails.

   The influence of higher-order dispersion on the dynamics of solitons was studied intensively, see
e.g. Refs.~\cite{Agra07,Blan16,Krug18,Tahe19,Tam19,Rung20,Melc20,Widj21,Li22,Akhm94,Karl94,Akhm95,Zakh98,Bian04,Tsoy07}.
Usually, two extreme cases are
investigated. Namely, either TOD and FOD are treated as perturbation to the GVD
effect~\cite{Akhm95,Zakh98,Bian04,Tsoy07}, or the FOD effect is considered as a dominant
one~\cite{Blan16,Krug18,Tahe19,Tam19,Rung20,Melc20,Widj21,Li22,Akhm94,Karl94}.
The former (latter) approach is valid far from
(close to) zero dispersion points (ZDPs). In particular, the consideration of a system as a medium with
pure-quartic dispersion is only valid near a specific ZDP, where both GVD and TOD are negligible.
In contrast to previous works, we make no assumptions on values of GVD, TOD, and FOD effects.
We show also that TOD does not result in the pulse asymmetry, if it acts together with FOD.
Our results, based on the variational approach, indicate clearly that the joint action of GVD, TOD,
and FOD can be balanced by Kerr nonlinearity, giving solitons with symmetric shapes.
Numerical simulations of Eq.~(\ref{gnlse}) support this conclusion.

   The dispersion relation of linear waves, $\psi(\tau, z) \sim \exp[i(\eta(\omega) z - \omega \tau)]$,
of Eq.~(\ref{gnlse}) at $\gamma = 0$ has the following form:
\begin{equation}
  \eta(\omega) = \frac{\beta_2}{2} \omega^2  + \frac{\beta_3}{6} \omega^3 +
    \frac{\beta_4}{24} \omega^4.
\label{drel}
\end{equation}
Then  $\eta_1(\omega)  \equiv d \eta(\omega) /d\omega =
\beta_2 \omega +  \beta_3 \omega^2 / 2 + \beta_4 \omega^3 /6$ and $\eta_2(\omega)
\equiv  d^2 \eta(\omega) /d\omega^2 =
\beta_2 + \beta_3 \omega + \beta_4 \omega^2 /2 $ are the inverse of
the group velocity and the second dispersion parameter of linear waves at
$\omega$, respectively.

  It is known~\cite{Agra07} that in media with GVD only, $\beta_3 = \beta_4 = 0$, bright solitons
do not exist when $\beta_2 \gamma > 0$. Following this relation for dispersive media, when all
$\beta_j$, $j = 2, 3$, and $4$,  are involved, one would expect that solitons do not exist
when $\eta_2(b) \gamma > 0$, where $b$ is the soliton frequency. Our theory gives different
conditions, see below.

  Equation~(\ref{gnlse}) has the following Lagrangian density:
\begin{eqnarray}
  {\cal L} &=& \frac{i}{2} \left( \psi^{*} \psi_{z} - \psi \psi_{z}^{*} \right) +
    \frac{\beta_2}{2} |\psi_{\tau}|^2 +
\nonumber \\
  && i \frac{\beta_3}{12} \left( \psi_{\tau}^{*} \psi_{\tau \tau} -
       \psi_{\tau} \psi_{\tau \tau}^{*} \right)  +
  \frac{\beta_4}{24} |\psi_{\tau \tau}|^2 + \frac{\gamma}{2} |\psi|^4,
\label{dens}
\end{eqnarray}
where the star means the complex conjugation.

   We use a trial function in the form of the Gaussian function:
\begin{equation}
  \psi(\tau, z) = A \exp[-(\tau - \tau_c)^2/ (2 a^2)] e^{i[\phi - b(\tau - \tau_c) + c (\tau - \tau_c)^2]}.
\label{trial}
\end{equation}
Here, parameters $A, a, \tau_c, b, c$, and $\phi$ are the soliton amplitude, width, position
of the center, linear phase parameter (the soliton frequency), chirp parameter, and phase
parameter, respectively. All these parameters are assumed to be functions of $z$. The minus
sign of a term proportional to $b$ is taken for convenience. The actual form of
solitons differs from Eq.~(\ref{trial}). In particular, a soliton can have oscillating
tails~\cite{Tam19,Akhm94}. However, numerical simulations show that trial
function~(\ref{trial}) captures well the overall soliton shape, so the parameter values
predicted are close to the actual ones.

   The Lagrangian $L = \int_{-\infty}^{\infty} {\cal L} d \tau$ is expressed in terms of the pulse
parameters, using trial function~(\ref{trial}). The Euler-Lagrange equations for $L$ give the
following equations:
\begin{eqnarray}
  a'&=& -c \left[2 \eta_2(b)\, a + \frac{\beta_4}{2} \left( a^{-1} + 4 c^2 a^3\right) \right],
\label{de_w} \\
  c'&=& \frac{1}{2} \eta_2(b) \left( 4 c^2 -  a^{-4} \right) -
    \frac{E_0 \gamma}{2 \sqrt{2 \pi}}\, a^{-3} +
\nonumber \\
   && \frac{\beta_4}{8}  \left( - a^{-6} + 16 c^4 a^2 \right),
\label{de_c} \\
  \tau_c'&=& \eta_1(b) +
    \frac{\beta_3 + \beta_4 b}{4} \left( a^{-2} + 4 c^2 a^2  \right),
\label{de_tc} \\
  \phi'&=& \frac{\beta_2}{2} \left( a^{-2} - b^2 \right) +
     \frac{ \beta_3}{12} \left( 3 b a^{-2} - 4 b^3  - 12 b c^2 a^2 \right) +
\nonumber \\
   && \frac{ \beta_4}{32} \left( 3 a^{-4} - 4 b^4  + 8 c^2 - 32 b^2 c^2 a^2 - 16 c^4 a^4 \right)  +
\nonumber \\
   &&  \frac{5 \gamma E_0}{4 \sqrt{2 \pi}\, a},
\label{de_phi}
\end{eqnarray}
and $b\,' = 0$, where the prime denotes $d / dz$. Parameter $E_0 =
\sqrt{\pi} A^2 a = \sqrt{\pi} A^2(0) a(0)$ does not depend on $z$, and represents the
initial energy of the pulse. Equations, similar to Eqs.~(\ref{de_w})-(\ref{de_phi}),
have been obtained previously, see e.g. Refs.\cite{Bian04,Tsoy07}. However, these equations were
used mainly to analyze the influence of higher order effects on the soliton of the unperturbed
NLS equation. Pure quartic solitons have been studied by the same method in Ref.~\cite{Li22}, but
only for zero soliton frequency. Here, we are interested in the existence of stationary solitons in
the presence of higher-order dispersion.

  Equations~(\ref{de_w}) and~(\ref{de_c}) constitute a closed set because their right-hand sides,
$f_a(a, b, c)$ and $f_c(a,b,c,E_0)$, do not depend on $\tau_c$ and $\phi$. The right-hand
sides of Eqs.~(\ref{de_tc}) and~(\ref{de_phi}) correspond to the soliton velocity $1/v$ in the
retarded frame and the phase coefficient $\delta$, respectively. Equations~(\ref{de_w})
and~(\ref{de_c}) have the invariant, which is the effective Hamiltonian of these equations:
\begin{eqnarray}
  H(a, c)= 16 \beta_4 a^4 c^4 + 8 c^2 (\beta_4 + 4 \eta_2(b)\, a^2) +
\nonumber \\
    32 \eta(b) + \beta_4 a^{-4} + 8 \eta_2(b)\, a^{-2} + \frac{16\gamma E_0}{\sqrt{2\pi}} a^{-1}.
\label{ham}
\end{eqnarray}
Using $H(a, c) = H(a(0), c(0))$, one can express variable $c$ in terms of $a$, and substitute it
into equation for $a'$, or $a''$. In the latter case, the equation for the soliton width
describes the motion of a particle with coordinate $a$ in an effective potential.

  Stationary solutions are found from conditions $f_a(a, b, c)= 0$ and $f_c(a,b,c,E_0) = 0$. From
Eqs.~(\ref{de_w}) and~(\ref{de_c}), it follows that stationary states exist only when $c = 0$.
Then, the stationary soliton width $a_s > 0$ is determined from the following equation:
\begin{equation}
   a^3 + s_1 a^2 + s_2 = 0,
\label{a_stat}
\end{equation}
where $s_1 = \sqrt{2\pi}\, \eta_2(b) /(\gamma E_0)$, and $s_2 = \sqrt{2\pi}\, \beta_4 /( 4\gamma E_0)$.
Applying the Sturm's theorem for the number of positive roots to Eq.~(\ref{a_stat}), we obtain the following
result:

   (i) If $(s_1 > 0$ and $s_2 >0)$, or $(s_1 < 0$ and $s_2 > s_{2,\mathrm{th}})$, then Eq.~(\ref{a_stat})
does not have positive roots, where $s_{2,\mathrm{th}} = - 4 s_1^3/ 27$.

  (ii) For any $s_1$, if $s_2 < 0$, then Eq.~(\ref{a_stat}) has one positive root.

  (iii) If $s_1 < 0$ and $0 < s_2 < s_{2,\mathrm{th}}$, then Eq.~(\ref{a_stat}) has
two positive roots.

The first condition of Case (i) is reduced to ($\eta_2(b) \gamma > 0$ and $\beta_4 \gamma > 0$), c.f.
with the standard NLS equation. The second condition of Case (i) indicates that solitons do not
exist also for negative $\eta_2(b) \gamma$ and corresponding $\beta_4$.
Solitons for parameters from Case (iii) are mostly non-stationary due to the interaction with linear waves,
see the corresponding discussion below. Also, notice that solitons exist for any sign of $\gamma$.

   Though Eq.~(\ref{a_stat}) can be solved analytically, this gives a complicated dependence of $a_s$ on the
system parameters. Therefore, it is useful to consider some limiting cases. Firstly, we consider the case
of small $\beta_4$, namely, if $\beta_4, \beta_4 b^2 a^2 \ll \hat{\beta} \equiv \beta_2 + \beta_3 b$,
then
\begin{equation}
  a_s \approx -  4 \hat{\beta} / p - (a^2 + 32 \hat{\beta}^2 b^2) \beta_4 / (16 \hat{\beta}^2 p),
\end{equation}
where  $p = 4 \gamma E_0 / \sqrt{2\pi}$.
The second case is for small $\beta_2, \beta_3$, and $b$, namely,
if $\beta_2 a^2, \beta_3 b a^2, \beta_4 b^2 a^2 \ll \beta_4$, then
\begin{equation}
  a_s \approx  (-\beta_4/p)^{1/3} - 4\eta_2(b) / (3p).
\end{equation}
Therefore, for large $|\beta_4|$, solitons exist when $\beta_4 \gamma < 0$.
Having root $a_s$ of Eq.~(\ref{a_stat}), the stationary amplitude is found as
$A_s = [E_0 / (\sqrt{\pi}\, a_s)]^{1/2}$. Then $(A_s, a_s, b)$ and $c = 0$, together with $1/v_s$
and $\delta_s$, correspond to stationary parameters of a GQS.

   The theory predicts that in absence of $\beta_3$ and $\beta_4$, the stationary soliton velocity $1/v_s$
coincide with the inverse of the group velocity $\eta_1(b)$ of linear waves. The inclusion of
higher-order dispersion breaks this relation, see Eq.~(\ref{de_tc}). In particular, even at the
extremum of the dispersion relation, at $b = 0$, we have solitons, moving due to $\beta_3$.
This result is supported by numerical simulations of Eq.~(\ref{gnlse}). A related
observation is that a static soliton with $1/v_s = 0$  can have a phase dependence on time,
$b \neq 0$. The difference of the soliton velocity $1/v_s$ from $\eta_1(b)$ can be used for
slow light and fast light applications of solitons.

  Equations~(\ref{de_w})-(\ref{de_phi}) describe the adiabatic
dynamics of a soliton. These equations do not take into account the interaction of the
soliton with linear waves. However, this interaction can be accounted for qualitatively, using
the following arguments. One can distinguish two different ways of generation of linear
waves by solitons. When an initial pulse differs slightly from the stationary profile, the
pulse adjusts its form to the stationary one, radiating the excess as linear
waves. This adjustment is observed as damped oscillations of the soliton width and
amplitude. Such a type of interaction is accounted for, to some extent, in
Eqs.~(\ref{de_w})-(\ref{de_phi}) by the inclusion of the chirp parameter $c$. Namely,
if the chirp is absent in trial function~(\ref{trial}), $c \equiv 0$, width $a$ is constant on $z$,
even when $a(0)$ is different from the stationary value. The variational approach with the
chirp included treats the interaction with linear waves as a modulation of the soliton
phase~\cite{Ande83}. In contrast to the actual dynamics, the method gives undamped oscillations
of the soliton shape, but predicts reasonably well the frequency.

   The second type is due to the resonance interaction of a soliton with linear waves~\cite{Akhm95}.
It occurs at frequencies where the soliton dispersion relation intersects with the
dispersion relation of linear waves. The soliton dispersion relation is
usually a straight line obtained from the following procedure. Let the
stationary soliton has the form $\psi_s(\tau, z) = A_s f(\tau - z/v_s) \exp[i(\delta_s z - b
\tau)]$, see Eq.~(\ref{trial}), where real $f(\tau)$ describes the soliton profile. Then, the
soliton spectrum $\Psi_s(\omega, z)$, obtained from the Fourier transform, is written as
$\Psi_s(\omega, z) = A_s F(\omega - b) \exp\{i[\delta_s + (\omega - b)/v_s]z\}$, where
$F(\omega)$ is the Fourier transform of $f(\tau)$. This expression indicates that the
soliton dispersion relation, or the dependence of the soliton propagation constant
$\eta_{\mathrm{sol}}(\omega)$ on frequency, is determined as the following
\begin{equation}
  \eta_{\mathrm{sol}}(\omega) = \delta_s + (\omega - b)/v_s.
\label{eta_sol}
\end{equation}
The linear dependence~(\ref{eta_sol}) means that a soliton propagates without dispersion,
$d^2 \eta_{\mathrm{sol}}(\omega) / d \omega^2 = 0$, since it is balanced by nonlinearity.
At frequencies $\omega_r$, defined by
\begin{equation}
  \eta_{\mathrm{sol}}(\omega_r) = \eta(\omega_r),
\label{match}
\end{equation}
resonance linear waves are generated due to the phase matching condition, see
Refs.~\cite{Akhm95,Zakh98,Bian04,Tsoy07}. The rate at which the soliton energy goes to
linear waves depends on values of the soliton spectrum at these resonant frequencies. The
arguments presented above are valid for media with an arbitrary order of dispersion.
Resonance condition~(\ref{match}) can be obtained rigorously from the analysis of small
modulations of the soliton, see e.g. Refs.~\cite{Akhm95,Bian04}.

   In absence of higher-order dispersion ($\beta_3 = \beta_4 = 0$), parameter
$1/v_s = \eta_1(b) = d\eta(b) /db$, see Eq.~(\ref{de_tc}). Therefore,
for media with quadratic dependence only,  $\eta_{\mathrm{sol}}(\omega)$  is a straight
line that is parallel to the tangent to the dispersion relation of linear waves at
frequency $b$, and shifted up by the amount depending on peak power $A_s^2$~\cite{Akhm95,Bian04,Tsoy07}.
In presence of higher-order terms ($\beta_3 \neq 0,  \beta_4 \neq 0$), the soliton velocity
$1/v_s$ differs from $\eta_1(b)$, therefore $\eta_{\mathrm{sol}}(\omega)$ is not
parallel to the tangent, see Eq.~(\ref{de_tc}).

   In media with quadratic and cubic dispersion terms ($\beta_4 = 0$), resonant linear waves are
always generated because $\eta_{\mathrm{sol}}(\omega)$ intersects $\eta(\omega)$. Though
the theory developed predicts the presence of solitons in media with cubic dispersion,
these solitons are not stationary due to the continuous transfer of energy from solitons to
linear waves. The lifetime of such solitons can be large if the resonant frequency is far
from center  $b$ of the soliton spectrum. In contrast, when the quartic term is included,
one can find a range of frequencies $b$, for which $\eta_{\mathrm{sol}}(\omega)$ does not
intersect with $\eta(\omega)$. For example, such frequencies can be found near the extrema
of $\eta(\omega)$.

   The discussion above can be generalized with the following statement.  Localized pulses are possible
in nonlinear media with any order of dispersion. If the highest-order dispersion term is
odd, then these pulses are non-stationary (quasi-stationary) due to the continuous
radiation of linear waves. If the highest-order dispersion term is even, stationary stable
pulses may exist. A necessary condition in the latter case is that the soliton dispersion
relation does not intersect with the dispersion relation of linear waves.

  We mention also about embedded solitons. The spectrum of these localized waves is located within the
spectrum of linear waves, see e.g. Refs.~\cite{Fuji97,Yang01}. Embedded solitons appear mainly in
multi-component systems~\cite{Yang01}, though they also exist in scalar systems with cubic-quintic
nonlinearity~\cite{Fuji97}. However, these solitons exist for particular relations of the
systems parameters. To the best of our knowledge, embedded solitons are not found for
a system with cubic nonlinearity only. Extensive numerical simulations of Eq.~(\ref{gnlse}) show
that when the intersection of spectra occurs (with the soliton parameters found
from the variational approach), no stationary solitons exist for $\beta_4 \gamma > 0$.

  To summarize, Eqs.~(\ref{de_w})-(\ref{de_phi}) are valid when $\eta_{\mathrm{sol}}(\omega)$ does
not intersect with $\eta(\omega)$.  As it follows from Eq.~(\ref{de_phi}), $\delta_s$ has
contribution $\sim \gamma E_0$. Line $\eta_{\mathrm{sol}}(\omega)$ is shifted up with the increase
of $E_0$ for $\gamma >0$. Then, for $\beta_4 > 0$ and large $E_0$,
$\eta_{\mathrm{sol}}(\omega)$ most likely intersects with $\eta(\omega)$.
Therefore, case $\beta_4 < 0$ is more favorable for the existence of stationary solitons, when $\gamma >0$.

\begin{figure}[ht]
\centerline{
  \includegraphics[width=4.6cm]{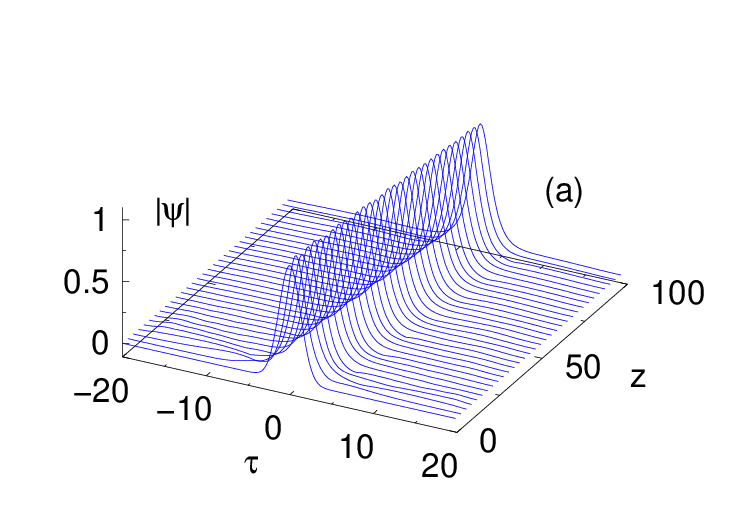}\hskip-0.5cm \includegraphics[width=4.6cm]{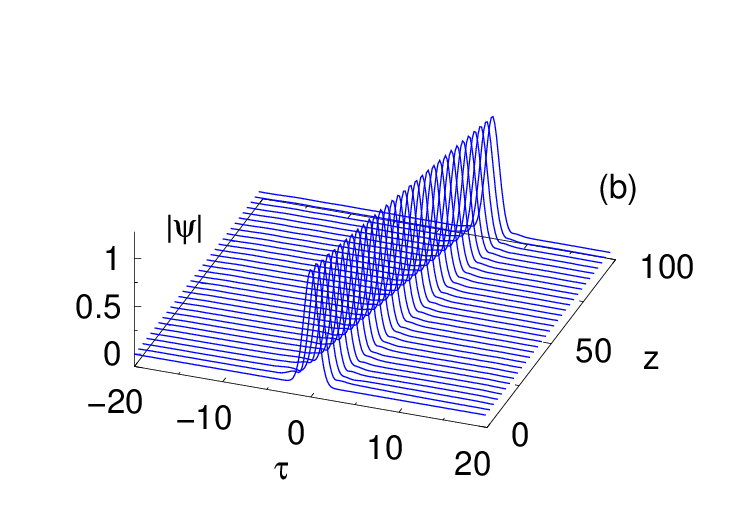} }
\centerline{
  \includegraphics[width=4.6cm]{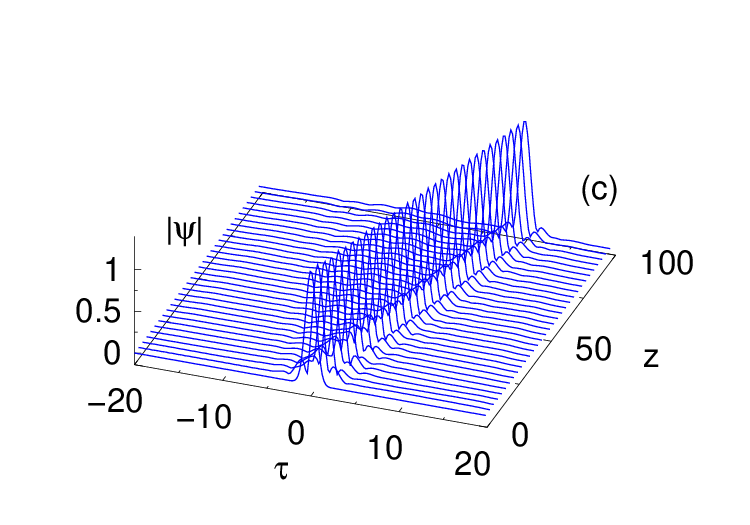}\hskip-0.5cm \includegraphics[width=4.0cm]{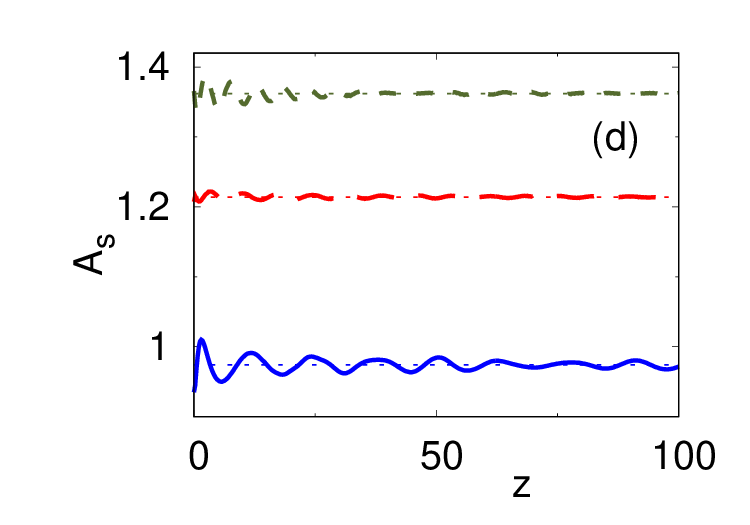} }
\caption{(a)-(c): The dynamics of solitons for (a) $(\beta_2, \beta_3, \beta_4)= (-1, 0.2,
-0.2)$, (b) $(-0.2, 0.2, -1)$, and (c) $(0.2, 0.2, -1)$. Other parameters are  $\gamma=1$
and $E_0 = 2$. (d) The dynamics of $A_s$ on $z$ for parameters from Fig.~\ref{fig:dyn}(a)
(solid line), Fig.~\ref{fig:dyn}(b) (long-dashed line), and Fig.~\ref{fig:dyn}(c) (short-dashed line).
Horizontal dotted lines correspond to values of stationary amplitudes predicted by the variational
approach.}
\label{fig:dyn}
\end{figure}

\section{Numerical simulations and transformations}
\label{sec:num}

   In order to check theoretical predictions, we perform numerical simulations of Eq.~(\ref{gnlse}).
For this purpose, we take all variables as dimensionless. We consider three cases: i) small
$|\beta_4|$, ii) small $|\beta_2|$, and iii) $\beta_2 > 0$. We take such values of
parameters that there is no intersection of $\eta_{\mathrm{sol}}(\omega)$ and
$\eta(\omega)$. The split-step Fourier method~\cite{Agra07} is used. The size of the
computational window is $T_{\mathrm{num}} = 30$-$50$, and the number of discretization points
is $512$-$1024$. Absorbing boundary conditions are used to prevent the reflection of
linear waves from edges. Initial conditions are in the form of Eq.~(\ref{trial}). A
relatively small value of $\beta_3 = 0.2$ is taken for convenience to restrict
the size of the computational window because $1/v$ grows with an increase of $\beta_3$,
see Eq.~(\ref{de_tc}). Theoretical predictions have the similar accuracy for larger $\beta_3$
as well.

   Figure~\ref{fig:dyn}(a)-(c) shows the dynamics of solitons for the three sets of parameters,
and $E_0 = 2$. Since initial profiles are approximate, solitons adjust their shapes,
emitting linear waves. In Fig.\ref{fig:dyn}(c), the field at large $z$ has oscillating
tails. Though, trial function~(\ref{trial}) is different from this form, the theory gives
acceptable values for stationary parameters with a deviation of 10-20\%, even for larger
values of $\beta_2\ (\ge 0.5)$. In Fig.~\ref{fig:dyn}(d), variations of the soliton amplitudes for the
dynamics in Fig.~\ref{fig:dyn}(a)-(c) are presented. The amplitudes tend to stationary
values via damped oscillations. Also, Fig.~\ref{fig:dyn} demonstrates the stability of
solitons to small modulations.

\begin{figure}[ht]
\centerline{ \includegraphics[width=4.4cm]{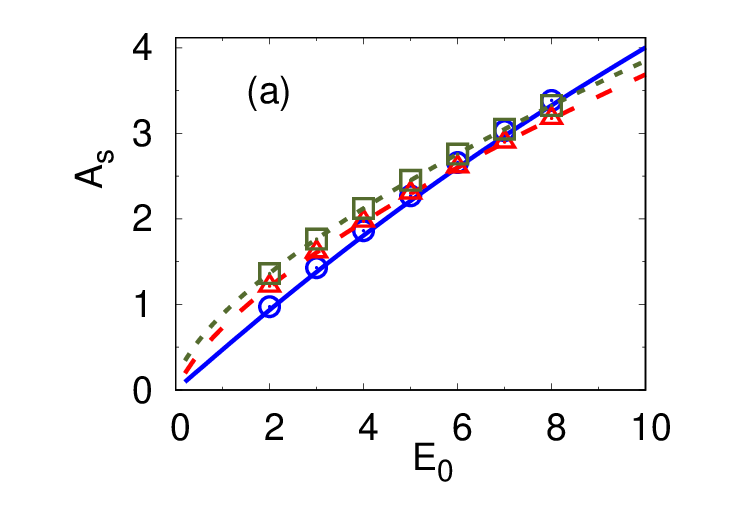}
   \includegraphics[width=4.4cm]{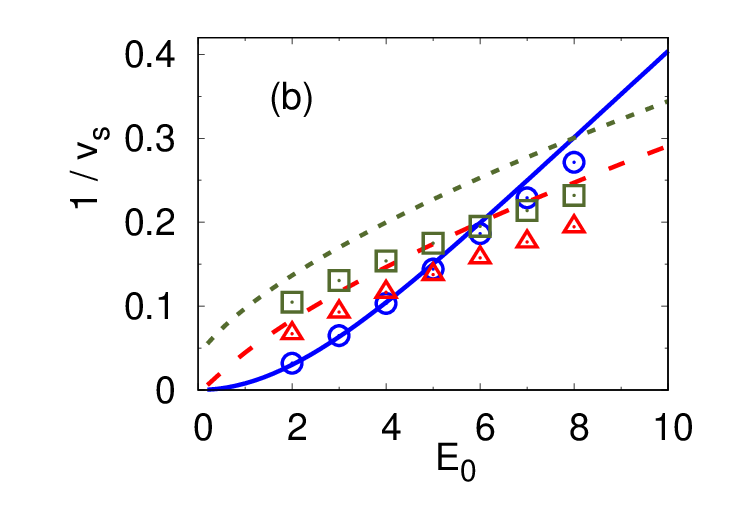} }
\centerline{ \includegraphics[width=4.4cm]{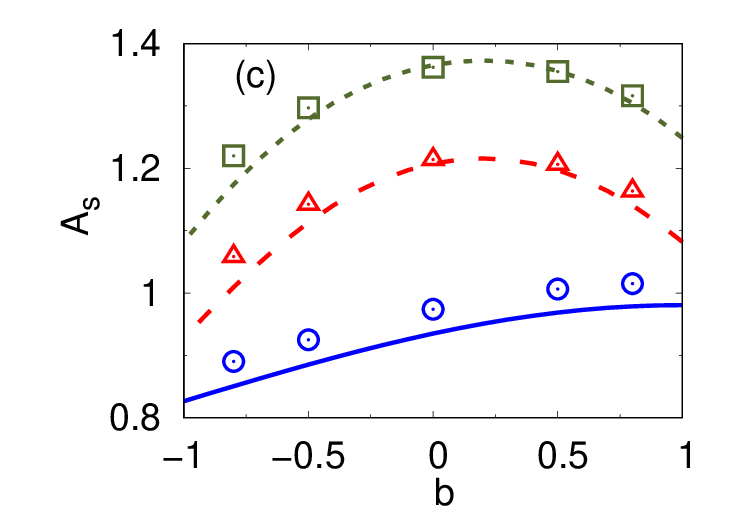}
   \includegraphics[width=4.4cm]{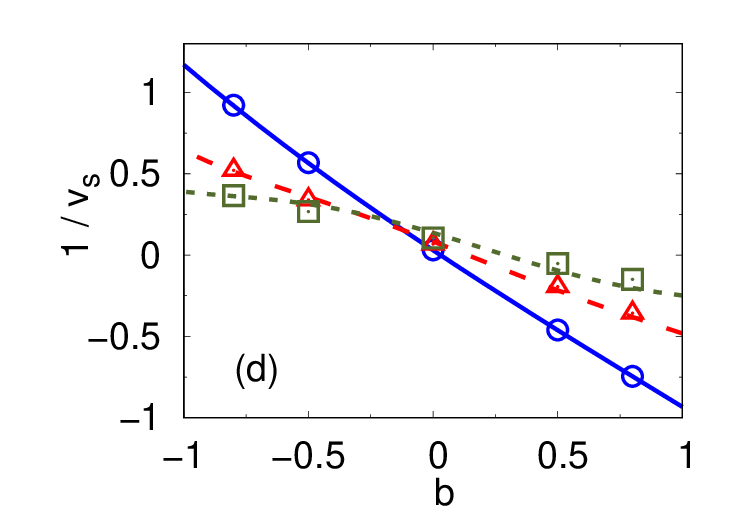} }
\caption{Soliton parameters as  functions of (a) and (b) $E_0$ at $b = 0$, and (c) and (d)
$b$ at $E_0 =2$. Solid lines (circles): $(\beta_2, \beta_3, \beta_4)= (-1, 0.2, -0.2)$,
dashed lines (triangles): $(-0.2, 0.2, -1)$, and dotted lines (squares): $(0.2, 0.2, -1)$.
Points corresponds to parameters found from numerical simulations of Eq.~(\ref{gnlse}). }
\label{fig:dep}
\end{figure}

  Dependencies of $A_s$ and $1/v_s$ on $E_0$ and $b$ are presented in Fig.~\ref{fig:dep}.
Soliton velocity $1/v_s$ is found as the average velocity over range $z \sim 20$-$50$ after the adjustment
process. There are small deviations of the predicted values from those found from numerical simulations.
However, the theory gives correctly the overall trend of all dependencies in Fig.~\ref{fig:dep}.
The soliton amplitude increases on $E_0$, and correspondingly, the soliton width $a_s$ decreases
on $E_0$. Contributions of the dispersion terms can be compared using the characteristic lengths~\cite{Agra07}
$L_{\mathrm{GVD}} = a_s^2 / |\beta_2|$, $L_{\mathrm{TOD}} = a_s^3 / |\beta_3|$, and
$L_{\mathrm{FOD}} = a_s^4 / |\beta_4|$. The smaller the length is, the more important is the
contribution of the corresponding effect.   Since $a_s$ varies on $E_0$ and $b$, relative contributions
of the dispersion terms are changed as well. For example, for $(\beta_2, \beta_3, \beta_4) = (-1, 0.2, -0.2)$,
the corresponding lengths are
$L_{\mathrm{GVD}} = 1.67, L_{\mathrm{TOD}} = 10.8$, and
$L_{\mathrm{FOD}} = 13.9$ at $E_0 = 2$, while at $E_0 = 10$,
$L_{\mathrm{GVD}} = 0.124, L_{\mathrm{TOD}} = 0.218$, and
$L_{\mathrm{FOD}} = 0.0767$. Therefore, as $E_0$ increases, the influence of TOD and FOD
increases as well.

   Ratios of $|\beta_2|$ and $|\beta_4|$, as in Figs.~\ref{fig:dyn} and~\ref{fig:dep}, can be
obtained in optical media at wavelengthes close to zero dispersion points. We consider,
as an example, the structure in Ref.~\cite{Blan16}. To find values of $\beta_3$ and
$\beta_4$, we fit the dependence $\beta_2(\lambda)$ in Fig.1(d) of Ref.~\cite{Blan16},
therefore obtained values of $\beta_3$ and $\beta_4$ can be slightly different from the
original values. At $\lambda = 1547.7\ \mathrm{nm}$, we get $\beta_2 = -2.7\ \mathrm{ps}^2
\mathrm{mm}^{-1}$,  $\beta_3 = -3.4\ \mathrm{ps}^3 \mathrm{mm}^{-1}$ $\beta_4 = -0.40\
\mathrm{ps}^4 \mathrm{mm}^{-1}$. Then $|\beta_2 / \beta_4|\, a_s^2 = L_{\mathrm{FOD}}
/L_{\mathrm{GVD}} \approx 5$, see Fig.~\ref{fig:dyn}(a), where $a_s = 0.85\ \mathrm{ps}$ is found from
Eq.~(\ref{a_stat}) for $E_0= 2\ \mathrm{pJ}$ and $\gamma = 4.1\cdot 10^3\
(\mathrm{W\,m})^{-1}$. At $\lambda = 1548.5\ \mathrm{nm}$, we get $\beta_2 = -0.79\
\mathrm{ps}^2 \mathrm{mm}^{-1}$,  $\beta_3 = -2.8\ \mathrm{ps}^3 \mathrm{mm}^{-1}$ $\beta_4 =
-1.5\ \mathrm{ps}^4 \mathrm{mm}^{-1}$. Then $|\beta_2 / \beta_4|\, a_s^2 \approx 0.2$,
see Fig.~\ref{fig:dyn}(b), where $a_s = 0.58\ \mathrm{ps}$ is found for the same $E_0$  and $\gamma$.

   The standard NLS equation, i.e. Eq.~(\ref{gnlse}) with $\beta_3 = \beta_4 = 0$, is invariant under
the Galilean transformation. It means that a moving solution of the NLS equation can be obtained from
a static solution by a corresponding change of variables. In contrast, the full Eq.~(\ref{gnlse})
is not Galilean invariant. The shape of the soliton can be altered as the velocity changes.
This property is ignored in trial function~(\ref{trial}). Nevertheless, this function gives a
reasonable approximation for solitons.

   It is possible to establish relations between exact solutions with different velocities of the
two related models.
For a particular choice of $\beta_2$, namely $\beta_2 = \beta_3^2 /(2\beta_4)$,
solutions of Eq.~(\ref{gnlse}) are related to solutions of the pure quartic NLS equation.
Let $\psi(\tau, z)$ be a solution of Eq.~(\ref{gnlse}), then $u(T, Z)$, defined from
$\psi(\tau, z)= [u(T, Z) / g^2] \exp[i (K Z - \Omega T)]$, is a solution of
the pure quartic NLS equation:
\begin{equation}
  i u_{\scriptstyle Z} +  \frac{\beta_4}{24} u_{TTTT} + \gamma |u|^2 u = 0,
\label{pure}
\end{equation}
where $T = \tau / g - z / (V g^4)$ and $Z= z/g^4$, $g$ is a free parameter, and
\begin{eqnarray}
  V &=& -  6 \beta_4^2 /(\beta_3^3 g^3),  \quad  \beta_2 = \beta_3^2  / (2 \beta_4),
\nonumber \\
  \Omega &=& -\beta_3 g /\beta_4, \quad K = -\beta_3^4 g^4/ (24\beta_4^3).
\end{eqnarray}
Alternatively, if $u(T, Z)$ is a solution of Eq.~(\ref{pure}), then $\psi(\tau, z)$, defined
from $u(T, Z) = [\psi(\tau, z) / g^2] \exp[i (K z - \Omega \tau)]$, is a solution of Eq.~(\ref{gnlse}),
provided that $\tau = T / g -  Z /(V g^4)$, $z = Z / g^4$, and
\begin{eqnarray}
  \beta_2 &=& \beta_4 \Omega^2 /2,  \quad \beta_3 =  \beta_4 \Omega,
\nonumber \\
  V &=& 6 / (\beta_4 \Omega^3),  \quad K = -\beta_4 \Omega^4/ 8,
\label{tr2}
\end{eqnarray}
where $g$ and $\Omega$ are free parameters.
Transformations~(\ref{tr2}) has been obtained also in Ref.~\cite{Widj21}.
Therefore, static and moving (in the retarded reference frame) solutions of Eq.~(\ref{gnlse}) can be
obtained from solutions of Eq.~(\ref{pure}) that move, in general, with different velocities, and
vice versa.

\section{Conclusions}
\label{sec:concl}

   In conclusion, we have demonstrated that stationary pulses, generic quartic solitons,
can propagate in media with GVD, TOD, and FOD. Numerical simulations of Eq.~(\ref{gnlse})
show that these pulses are stable for sufficiently long distances. Conditions in terms of
the system parameters have been identified for the existence of GQS. In particular, these
solitons exist both for the positive GVD and negative GVD parameters.  Parameters of
stationary solitons for different energies and soliton frequencies have been found
approximately. Values of these parameters are close to those found numerically. It has been
demonstrated that the soliton velocity in general quartic media differs, in principle, from
the inverse of the group velocity of linear waves. It has been shown that the resonance
interaction of a pulse with linear waves can prevent the existence of stationary solitons.
Transformations, that connect solutions of Eq.~(\ref{gnlse}) with those of
Eq.~(\ref{pure}), have been obtained. Our analysis provides strong support for a conjecture
that stable solitons can exist in media with a general form of dispersion if the
highest-order dispersion term is even.

  Our results suggest an alternative view on the dynamics of pulses in dispersive nonlinear
media, in particular, during supercontinuum generation. Usually, the dynamics is considered
as a perturbation of solitons of the standard (with GVD only) NLS model. However, the dynamics
can also be treated as an adjustment of pulses to stationary solitons associated with
higher-order dispersion.

\begin{acknowledgments}
 The work was supported by the Ministry of Higher Education, Science and Innovation
of the Republic of Uzbekistan.
\end{acknowledgments}

\end{document}